\documentclass[showpacs,aps,prb,lettersize,12pt]{article}

\usepackage{amsmath}
\usepackage{graphicx}
\usepackage{appendix}
\usepackage{authblk}
\usepackage{txfonts} 

\begin{document}
\title{A simple analytical approach to describe disease spread on a network}
\author[1]{Bahman Davoudi}
\author[2]{Fred Brauer}
\author[1,3,*]{Babak Pourbohloul}

\affil[1]{Division of Mathematical  Modeling, British Columbia Centre
for Disease Control, Vancouver, British Columbia, Canada}
\affil[2]{Department of Mathematics, University of British Columbia,
Vancouver, British Columbia, Canada}
\affil[3]{School of Population \& Public Health, University of British
Columbia, Vancouver, British Columbia, Canada}
\affil[*]{Corresponding author}
\maketitle
\setlength{\baselineskip}{1.5\baselineskip} 

\begin{abstract}
We investigate the time evolution of disease spread on a network by
using the concept of generations. We derive a set of equations, which
can be used to determine the average epidemic size. 
We find a very good agreement between the analytical and simulation
results. The details of approximations and the possibility of generalization
or improvement are discussed. 
\end{abstract}

\section{Introduction}

The spread of disease in a population has been the subject of extensive
studies recently. This is an important field of investigation as it
allows us to plan a better strategy for intervention in order to reduce
the burden of an epidemic. This is, of course, an important issue
with regard to the health of the populations of humans and other species.

Disease spread has been studied by a variety of different models and
approaches. There are three basic types of model, namely, compartmental
models \cite{1,2,2a,3}, network models \cite{4,5,6,7,8,9}, and agent-based
models \cite{10,11,12}. Each type of model approaches the problem
from a different angle and has its own level of simplifications or
complexities, advantages or disadvantages. There has been a long debate
on the question of which type of model can be considered as a parsimonious
model to address the spread of disease (or a class of diseases) in
an adequate manner. Among the above models, the compartmental model
makes a full mixing assumption for the contact between individuals.
The full-mixing assumption, which states that an infected individual
is equally likely to spread the disease to any susceptible individual,
is the basic assumption for the compartmental model, although it is
possible to consider compartments with different activity levels.
However, compartmental models may not be suitable for describing the
initial phase of a disease outbreak when the number of infectious
individuals is small and the progression of an outbreak is dominated
by stochastic fluctuations. In network models, each infected individuals
can transmit the disease to those individuals to which they are connected.
A connection (also known as edge or link) between two individuals
exists if disease transmission between them is possible. The edge
is a static object in a static network model and a dynamic object
in a dynamic network model or an agent-based model. To address several
public health questions of import, static network models seem to be
appropriate candidates to represent human interactions so long as
the exact timing of transmission between two individuals is not known
and that every individual's connections over the period of one disease
generation time is well represented in the network. A disease generation
time is roughly equal to the incubation period plus the infectivity
period of that disease. This, in fact, is the basis of contact tracing
approaches in public health outbreak management in which all contacts
within one disease generation are identified and tested for potential
transmission. In should be noted that disease generation time is different
from the generation interval, the latter being the time between when
an individual is infected and the time that he/she infects someone
else.

Disease spread on a network has been studied by several approaches,
including the mean value reaction equation \cite{14,15,16,17,18},
the generating function formalism \cite{19,20}, the time evolution
of the generating function \cite{21,22}, and computer simulations
\cite{23,24}. Among these, the second approach focuses on only the
final size value of an epidemic while the others study the time evolution
of the disease spread.

In the following section we first present the theory behind the current
calculation. This requires a short introduction to networks, the concept
of generations and then the definition of different types of links
and vertices. We next obtain a set of equations allowing us to find
the number of different links and vertices in a given generation.
We present the numerical result of this approach in the subsequent
section in order to validate the approximations used in our calculations
as well as to study other aspects of the disease spread on the network.
Finally, we summarize the results and provide our overview on the
subject, including the possibility of extensions and improvements.

\section{Theory}

We study the evolution of the disease spread on a network by using
the notion of the generation. A network is built of vertices (nodes)
representing individuals and links (edges) representing contacts between
them that could potentially lead to disease transmission. We call
two vertices ``neighbors'' if they are connected by a link. The
degree of a vertex is defined as the number of links connecting the
vertex to its neighbors, and the probability distribution of these
degrees over the whole network is called the network's degree distribution. A
network is completely defined when we know the number of vertices
and the way in which they are connected to one another. In most practical
situations however, for instance, when the population is large, it
is hard and not even necessary to keep track of the connectivity of
vertices. One can create a network which is suitable for studying
many aspect of the disease spread by respecting a minimum of requirements.
The minimal requirement is to create a network that satisfies the
desired degree distribution. The degree of each vertex can be assigned
randomly and the vertices can be connected randomly as well provided
that this minimum requirement is met; the connections in the real population
can differ from this otherwise completely random network due to potential
correlations between individuals, such as partnerships, friendships,
etc. This is measured by the clustering coefficient and can be included
in the network \cite{25,26}.

We divide the vertices into three classes, namely susceptible, infectious,
and removed classes (where removed means infectious in the past but
are no longer infectious). Such models are called SIR models to indicate
the three classes. An infectious vertex could infect each of its neighbors,
with a probability $T$ during the entire period $\tau$ of its infectivity.
In reality the infectivity is a complex function of time, depending
on the type of infection, and the infectivity and susceptibility could
also depend on the individuals. Here we model the infectivity by $T(t)=T\delta\left(t-\tau\right)$,
which means that total force of infection is applied at time $\tau$,
and we then remove the infective vertex. Normally one or more initially
infected individual brings an infection to a population but here for
the sake of simplicity we start our calculation with one initial infection.
The first infected individual can transmit the disease to a fraction
of its neighbors at time $\tau$ (the first generation). The first
infected individual then belongs to the removed class and the new
infective individuals are in the infective class. These new infective
individuals could transmit the infection to their neighbors after
a passage of time $\tau$ (the second generation). It is easy now
to recognize that the spread of the infection in the network follows
the concept of the generations. We have a number of infective individuals
at each given generation (period of $\tau$) and these individuals
are the carriers of the disease in the network in that generation.
The rate of new infections at each given generation for a random network
depends on the transmissibility \emph{T}, the degree distribution
within each class, the number of links that connect vertices in different
classes and the relative populations of vertices in each classes.
In the following we assume that the rate of new infections depends
on the average degree of each class rather than their degree distributions
and this leads to a significant simplification (we explicitly assume
that all vertices within each class in a given generation have the
same degree). Other generalizations will be discussed later.

We define $N_{r}^{g}$, $N_{i}^{g}$, $N_{s}^{g}$ as the number of
removed, infective and susceptible individuals in a given generation
$g$, respectively, and $z_{r}^{g}$, $z_{i}^{g}$, $z_{s}^{g}$ as
their average degrees. Each vertex in a given class is connected to
another vertex in the same or different class by an intra-link or
inter-link, respectively. The total number of intra-link connections
are specified as $L_{s}^{g}$, $L_{i}^{g}$, $L_{r}^{g}$ for links
within the susceptible, infectious and removed classes, respectively;
while the total number of inter-link connections between classes \emph{x}
and \emph{y} are shown as $L_{xy}^{g}$, where \emph{x} and \emph{y}
stand for the indices \emph{s}, \emph{i} and \emph{r}, representing
the three classes described earlier. Note that since $L_{xy}^{g}$
is the \emph{total} number of links between classes \emph{x} and \emph{y},
then $L_{xy}^{g}=L_{yx}^{g}$. Finally, $N$, $z$ and $L$ are the
total number of vertices, the average degree and the total number
of links of the whole network, respectively. Also, we will use the
subscript $\alpha$ as a generic index for \textit{r, i, s}.

Because we are assuming that infectious individuals are removed in
each generation, the number of infectious individuals in a generation
is exactly the number of new infections in that generation, and this
depends on the number of links between infectious and susceptible
individuals. Each link between two vertices can be constructed from
two stubs associated to the two vertices. A stub can be thought of
the segment of a link emanating from each vertex; therefore, the number
of stubs at a vertex is the degree of that vertex. The total number
of stubs in a given compartment $\alpha$ is given by $N_{\alpha}^{g}z_{\alpha}^{g}$
of which $2L_{\alpha}^{g}$ contribute to intra-links and the rest
contribute to the inter-links. It is then straightforward to establish
these identities:

\begin{equation}
\begin{array}{l}
{z_{r}^{g}N_{r}^{g}=2L_{r}^{g}+L_{rs}^{g}+L_{ri}^{g}}\\
{z_{i}^{g}N_{i}^{g}=2L_{i}^{g}+L_{is}^{g}+L_{ri}^{g}}\\
{z_{s}^{g}N_{s}^{g}=2L_{s}^{g}+L_{rs}^{g}+L_{is}^{g}}\end{array}\label{1}\end{equation}

The above set of equations imposes a strong constraint on the number
of intra- and inter-links. This set of equations allows us to find
the number of inter-links in terms of the number of intra-links, the
class average degree and the number of vertices in each class. The
number of inter-links is important for measuring the spread of the
disease in the network because an inter-link between the susceptible
and infective classes may lead to a new case of infection.

The average number of intra-links for each class can be estimated
in terms of $\{N_{\alpha}^{g},z_{\alpha}^{g}\}$ as follows. We consider
a stub of a specific vertex in a compartment $\alpha$ and calculate
the probability that this stub attaches to a stub corresponding to
another vertex within the same class. The total number of stubs to
which it could attach is \begin{equation}
Nz-z_{\alpha}^{g}.\end{equation}
 This means that this stub could in principle attach to any other
stubs in the whole network except the ones belonging to the same vertex,
i.e., self-loops are not allowed. The total number of stubs to which
this stub could attach in its own class is \begin{equation}
(N_{\alpha}^{g}-1)z_{\alpha}^{g},\end{equation}
 excluding self-loops thus the probability that we searched for is
given by \begin{equation}
\frac{(N_{\alpha}^{g}-1)z_{\alpha}^{g}}{Nz-z_{\alpha}^{g}}.\end{equation}
 This probability is in fact the average number of stubs attaching
to a stub in the same class. The probability that the second stub
of the given vertex attaches a stub of another vertex (excluding the
stubs of the last two vertices) is then given by \begin{equation}
\frac{(N_{\alpha}^{g}-2)z_{\alpha}^{g}}{Nz-2z_{\alpha}^{g}}\end{equation}
 and for the $j$th stub by \begin{equation}
\frac{(N_{\alpha}^{g}-j)z_{\alpha}^{g}}{Nz-jz_{\alpha}^{g}}.\end{equation}
 This process can be continued until all stubs of the given vertex
are exhausted or until all other vertices within the same class are
chosen, or more precisely while \begin{equation}
j\le\gamma_{\alpha}^{g}=\min(N_{\alpha}^{g},z_{\alpha}^{g}).\end{equation}
 The average number of intra-links associated to the given vertex
is the sum over all of the above terms. The average number of intra-links
is then given by \begin{equation}
\bar{L}_{\alpha}^{g}=\frac{N_{\alpha}^{g}}{2}\left[\gamma_{\alpha}^{g}-\left(\frac{Nz}{z_{\alpha}^{g}}-N_{\alpha}^{g}\right)\left(\psi^{(1)}\left(\frac{Nz}{z_{\alpha}^{g}}\right)-\psi^{(1)}\left(\frac{Nz}{z_{\alpha}^{g}}-\gamma_{\alpha}^{g}\right)\right)\right]\label{2}\end{equation}
 where $\psi^{(1)}(x)$ is the first polygamma function defined as
\begin{equation}
\psi^{(n)}(x)=(-1)^{n+1}n!\sum_{j=0}^{\infty}\frac{1}{(x+j)^{n+1}}.
\end{equation}
Equation (\ref{2}) leads to $\bar{L}_{\alpha }^{g} \approx\frac{z_{\alpha }^{g}(N_{\alpha }^{g})^{2}}{2Nz}$ in the limit $N\gg\gamma_{\alpha}^{g}$ and when $N_{\alpha}^{g}>z_{\alpha}^{g}$.

 We define $\tilde{N}_{\alpha}^{g}$ to be the number of vertices
in the complement of the $\alpha$ class and $\tilde{z}_{\alpha}^{g}$
as their average degree. The average number of intra-links for the
$\alpha$ class can also be calculated by the total average number
of its inter-links (the second term in the following equation) as
\begin{equation}
\bar{L}_{\alpha}^{g}=\frac{1}{2}\left[N_{\alpha}^{g}z_{\alpha}^{g}-N_{\alpha}^{g}\left[\tilde{\gamma}_{\alpha}^{g}-\left(\frac{Nz}{\tilde{z}_{\alpha}^{g}}-\tilde{N}_{\alpha}^{g}-1\right)\left(\psi^{(1)}\left(\frac{Nz}{\tilde{z}_{\alpha}^{g}}\right)-\psi^{(1)}\left(\frac{Nz}{\tilde{z}_{\alpha}^{g}}-\tilde{\gamma}_{\alpha}^{g}\right)\right)\right]\right],\label{3}\end{equation}
 where $\tilde{\gamma}_{\alpha}^{g}=\min(\tilde{N}_{\alpha}^{g},\tilde{z}_{\alpha}^{g})$.
The average number of intra-links can be evaluated by using \eqref{1}
and \eqref{2} {[}or \eqref{3}{]}.

Using $\bar{L}_{is}^{g}$and $\bar{L}_{rs}^{g}$ we can obtain the
average number of new infections in each generation. However, we must
take into account the \emph{finite size} effect, because in a network
of finite size we must take account of the possibility that two infectious
individuals may be linked to the same susceptible individual. In fact,
the number of new infections in the simplest scenario depends on $\bar{L}_{is}^{g}$.
We define the average number of transmitting and non-transmitting
links from the infectious compartment to be $\bar{\lambda}^{g}=T\bar{L}_{is}^{g}$
and $\bar{\lambdabar}^{g}=(1-T)\bar{L}_{is}^{g}$, respectively. The
number of new infections can be obtained by assigning the transmitting
links to the susceptible vertices. The first link makes a contribution
of one to the new infections. The probability that the second link
is attached to another vertex is roughly proportional to \begin{equation}
\frac{(N_{s}^{g}-1)z_{s}^{g}}{N_{s}^{g}z_{s}^{g}}=\frac{N_{s}^{g}-1}{N_{s}^{g}},\end{equation}
 and this in fact is the contribution of the second link to the average
number of new infections. We assume that all susceptible vertices
have the same degree, which is a good approximation when the degree
distribution is narrow; we discuss the more general case later. The
contribution of the $j$th link is then given by \begin{equation}
\frac{N_{s}^{g}-\bar{n}_{j-1}}{N_{s}^{g}},\end{equation}
 where $\bar{n}_{j-1}$ is the average number of new infections after
assigning $(j-1)$ links. The average number of the new infections
after assigning the $j$th links is \begin{equation}
\bar{n}_{j}=\bar{n}_{j-1}+\frac{N_{s}^{g}-\bar{n}_{j-1}}{N_{s}^{g}}.\end{equation}
 The number of new infections after assigning all transmitting links is
then given by \begin{equation}
\bar{N}_{i}^{g+1}=\frac{1-\beta_{g}^{\bar{\lambda}^{g}}}{1-\beta_{g}},\label{4}\end{equation}
 where $\beta_{g}=1-\frac{1}{N_{s}^{g}}$. It is straightforward to
show that \begin{equation}
\bar{N}_{i}^{g+1}=\bar{\lambda}^{g}-\frac{\bar{\lambda}^{g}\left(\bar{\lambda}^{g}-1\right)}{2N_{s}^{g}}\end{equation}
 where $N_{s}^{g}\gg\bar{\lambda}^{g}$, which means that the average
number of new infections is approximately equal to the number of transmitting
links in such a situation and the correction to this is given by $\frac{(\bar{\lambda}^{g})^{2}}{2N_{s}^{g}}.$
We point out that in deriving the above equation we ignore the requirement
that the infectious links emitted from one specific vertex cannot
connect to the same susceptible vertex (repeated links). We may take
this effect into account but the resulting formula does not show any
significant difference.

It is possible to extend the above equation to the case that the degree
distribution is very asymmetric with respect to the average degree
\begin{equation}
\bar{N}_{i}^{g+1}=\sum_{k}\bar{N}_{i}^{g+1}(k),\label{5}\end{equation}
 where \begin{equation}
\bar{N}_{i}^{g+1}(k)=\left\{ \begin{array}{ccc}
{\frac{1-\beta_{g}(k)^{\bar{\lambda}^{g}(k)}}{1-\beta_{g}(k)}} & {,} & {\bar{\lambda}^{g}(k)>1}\\
{\bar{\lambda}^{g}(k)} & {,} & {\bar{\lambda}^{g}(k)<1}\end{array}\right\} ,\end{equation}
 $\beta_{g}(k)=1-\frac{1}{N_{s}^{g}p_{s}^{g}(k)}$ and $\bar{\lambda}^{g}(k)=\frac{\bar{\lambda}^{g}kp_{s}^{g}(k)}{z_{s}^{g}}$.

An estimate for the average degree of each class completes the set
of equations which are required for the time evolution of the disease
on the network. This is similar to our earlier calculations presented
in Noel \textit{et al.} {[}18{]}. Because vertices with high degree
are more likely to receive infection during an epidemic process, we
would expect the average degree of the susceptible class to be less
than the average degree of the infective and removed class, and this
is of importance in describing the time evolution of the disease.

The first infection is chosen randomly so that the probability that
it has degree $k$ is given by $\Delta\tilde{p}_{1}(k)=p(k)$ where
$p(k)$ is the degree distribution of the network. We consider the
first chosen vertex as being removed. The new degree distribution
of the network remaining after removing the first vertex is $p_{1}(k)=p(k)$
because of the choice made for the first vertex. Now we pick the second
vertex by choosing a random stub. The new vertex could be one of the
neighbors of the first vertex in one of the networks, which is constructed
by knowing the degree distribution. The probability that the second
vertex has degree $k$ is given by $\Delta\tilde{p}_{2}(k)=\frac{p_{1}(k)k}{z_{1}}$.
This means that the degree distribution of removed vertices is $\tilde{p}_{2}(k)=\frac{\Delta\tilde{p}_{1}(k)+\Delta\tilde{p}_{2}(k)}{2}$
and so the degree distribution of the remaining vertices can be written
as $(N-2)p_{2}(k)=Np(k)-2\tilde{p}_{2}(k)$. The degree distribution
of the $j$th chosen vertex is $\Delta\tilde{p}_{j}(k)=\frac{p_{j}(k)k}{z_{j}}$
and in the same manner, the degree distribution of the removed and
remaining vertices are given by \begin{equation}
\begin{array}{l}
{\tilde{p}_{j}(k)=\frac{\sum_{l=1}^{j}\Delta\tilde{p}_{l}(k)}{j}}\\
{p_{j}(k)=\frac{Np(k)-j\tilde{p}_{j}(k)}{N-j}}.\end{array}\end{equation}
 The average degree of the removed class is $\bar{z}_{r}^{g}=\tilde{z}_{j}=\sum_{k}\tilde{p}_{j}(k)k$
when $j=\bar{N}_{r}^{g}$. The average degree of the infective class
is $\bar{z}_{i}^{g}=\frac{\left[(j+j')\tilde{z}_{j+j'}-j\tilde{z}_{j}\right]}{j'}$
when $j=\bar{N}_{r}^{g}$and $j'=\bar{N}_{i}^{g}$. Finally the average
degree of the susceptible class is $\bar{z}_{s}^{g}=z_{j}=\sum_{k}p_{j}(k)k$
when $j=\bar{N}_{s}^{g}$. We point out that in the derivation of
the above result, we assume that all vertices are connected to each
other and this might not be true for a general network.

Now we are in a position to estimate the effect of non-transmitting
links on the reduction of $N_{s}^{g}$. We call the total number of
non-transmitting links $\Lambda^{g}=\bar{\lambdabar}^{g}+L_{rs}^{g}$
and the degree distribution of susceptible $P_{0}^{g}(k)=p_{s}^{g}(k)$.
Assigning the first non-transmitting link without removing the vertex
changes the degree distribution to \begin{equation}
N_{s}^{g}P_{1}^{g}(k)=N_{s}^{g}P_{0}^{g}(k)+\frac{P_{0}^{g}(k+1)(k+1)}{Z_{0}^{g}}-\frac{P_{0}^{g}(k)k}{Z_{0}^{g}}.\end{equation}
 The second and third terms in the above equation are the contribution
of moving a vertex from $k+1\to k$ and $k\to k-1$ because of removing
one stub. It is straightforward to show that after assigning all links,
\begin{equation}
N_{s}^{g}P_{\Lambda^{g}}^{g}(k)=N_{s}^{g}P_{0}^{g}(k)+\sum_{j=0}^{\Lambda^{g}-1}\frac{P_{j}^{g}(k+1)(k+1)}{Z_{j}^{g}}-\frac{P_{j}^{g}(k)k}{Z_{j}^{g}}\label{6}\end{equation}
 is the effective degree distribution of the susceptible class after
assigning all non-transmitting links. The total number of degree zero
vertices is then simply given by $N_{s}^{g}P_{\Lambda^{g}}^{g}(0)$,
which should be subtracted from $N_{s}^{g}$ in \eqref{4}. One can
either use the effective degree distribution in \eqref{5} to take
the effect of non-transmissible links into account. Note that the
total number of degree zero vertices is not the only contribution
to the reduction of the number of susceptible vertices, which can
be reached by transmitting links. Small components make a very important
contribution especially if the average degree is low or if the degree
distribution has significant value in the low degree region.

\section{Numerical results}

Here we present our numerical results, which are mainly the solution
of \eqref{1} - \eqref{6} together with the equations for the average
degree of the classes. The first part of our numerical results focuses
on the justification of the approximations introduced in the previous
sections. We compare our analytical results against simulations for
different types of networks to ensure their validity. We examine our
analytical results for two distributions, namely Poisson $p(k)=\frac{z^{k}e^{-z}}{k!}$,
and exponential $p(k)\propto e^{\kappa k}$ (with $z=10$ and $\kappa=10$
for all figures).

The Poisson distribution corresponds to homogeneous mixing and is
in a sense the easiest example because the network is connected and
the probability of a small outbreak is small. In an exponential distribution,
there are many vertices of small degree and there is a higher probability
that the graph will include small components which would lead to small
disease outbreaks. Since we wish to describe the evolution of disease
spread for large epidemics, we should discard the simulations that
produce only small outbreaks. Keeping the small outbreak simulations
would lower the disease curve later in the progress of the epidemic.
In the figures of this section, we have incorporated the correction
of discarding the simulations giving only small outbreaks.

\begin{figure}[htb]
\centering{}\includegraphics[scale=0.35]{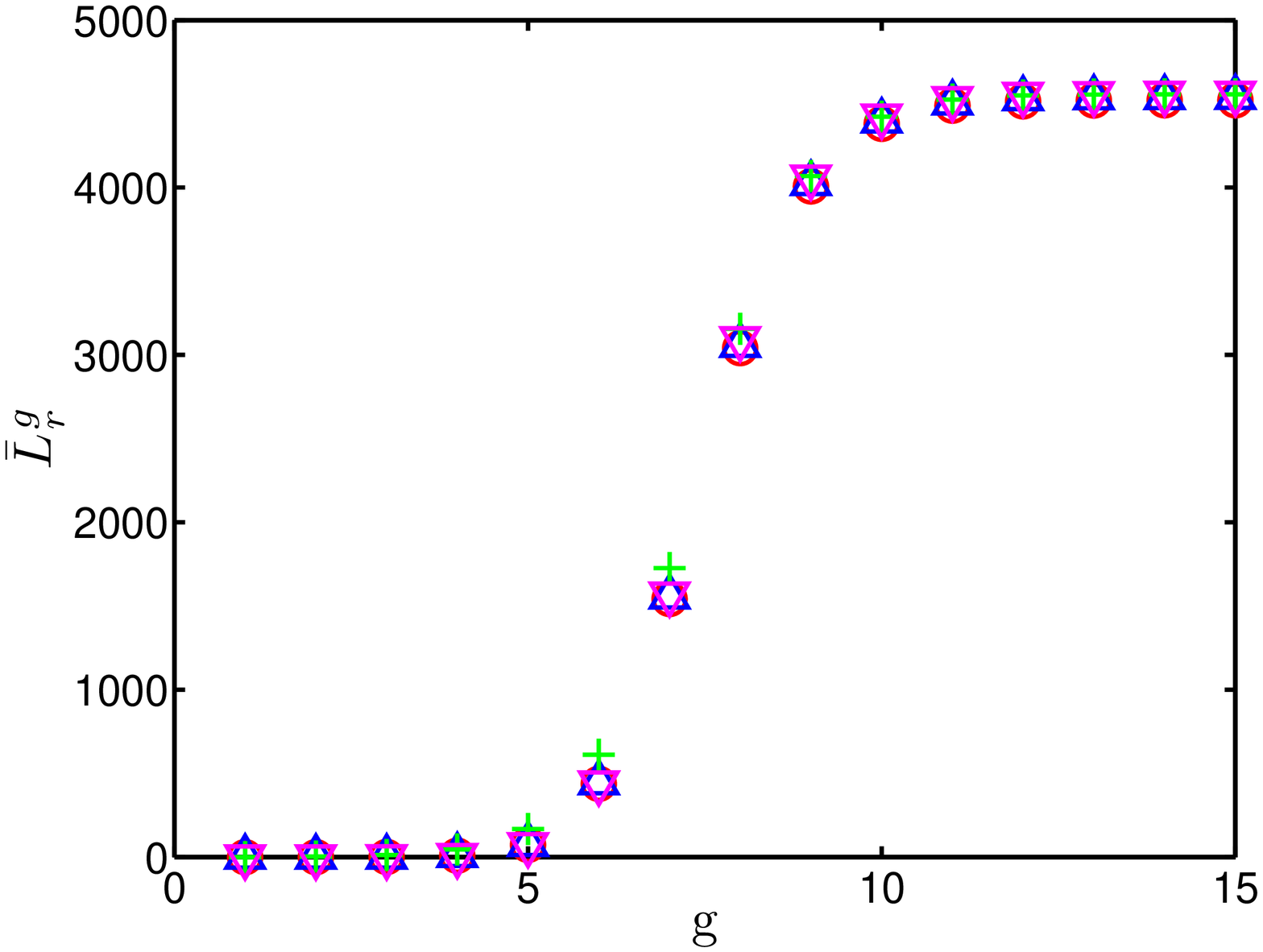} \includegraphics[scale=0.35]{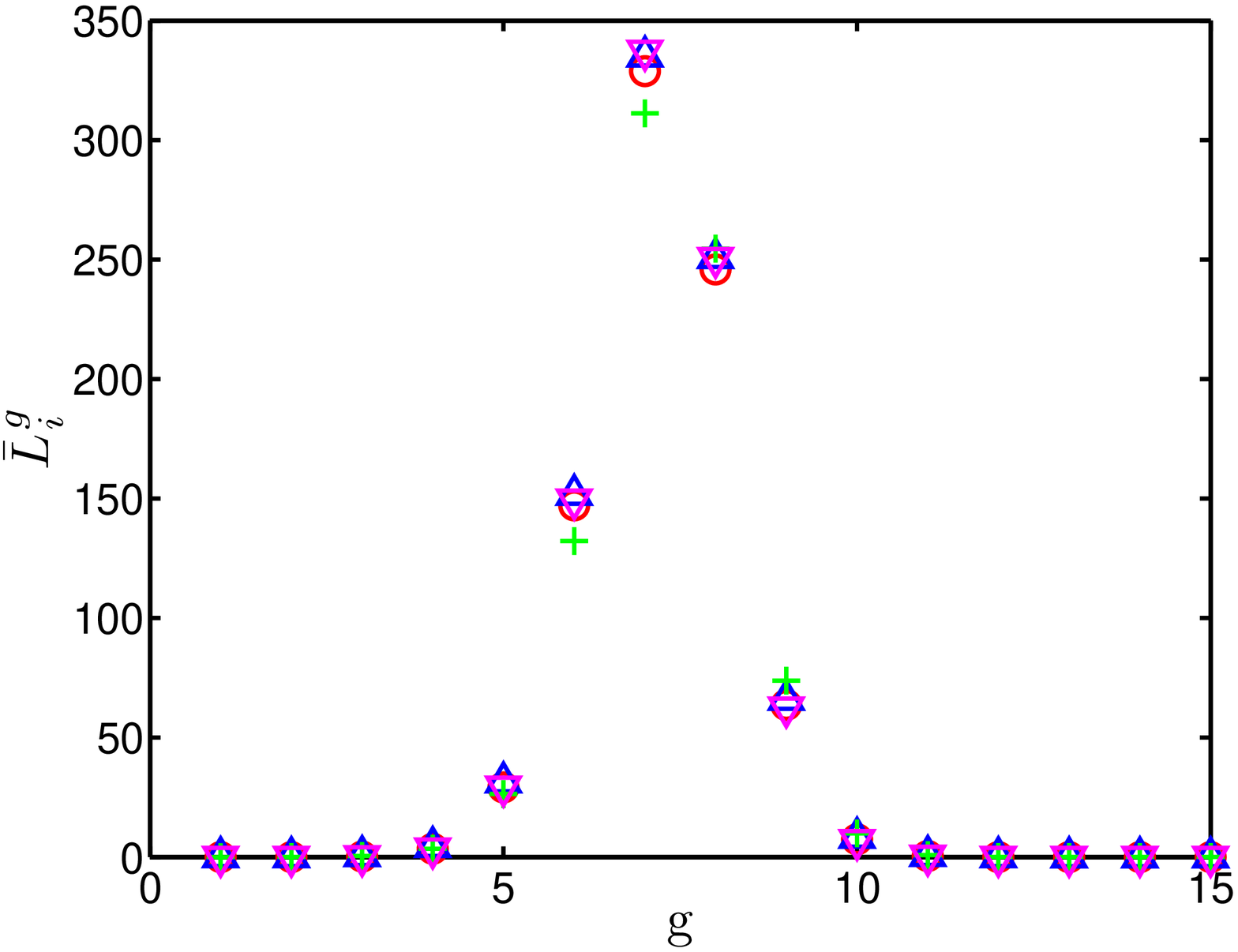}
\caption{The average number of intra-link for the removed (left panel) and
infectious (right panel) classes in terms of generations. The results
are for the Poisson distribution, $T=0.3$ and $N=1000$.\label{fig1}}

\end{figure}

Figure \ref{fig1} shows the average number of intra-links for the
removed (left panel) and infectious (right panel) classes in terms
of generations. The results are for the Poisson distribution, $T=0.3$
and $N=1000$. The plus signs show the direct outcome of simulation.
The circles and triangles are the outputs of \eqref{2} and \eqref{3}
respectively, using the number of vertices and average degree as simulation
inputs. The diamonds are the result of \eqref{2} using only the number
of vertices given by simulation as input. The average degree is calculated
by the analytical formula of the previous section.

\begin{figure}[htb]
\centering{}\includegraphics[scale=0.35]{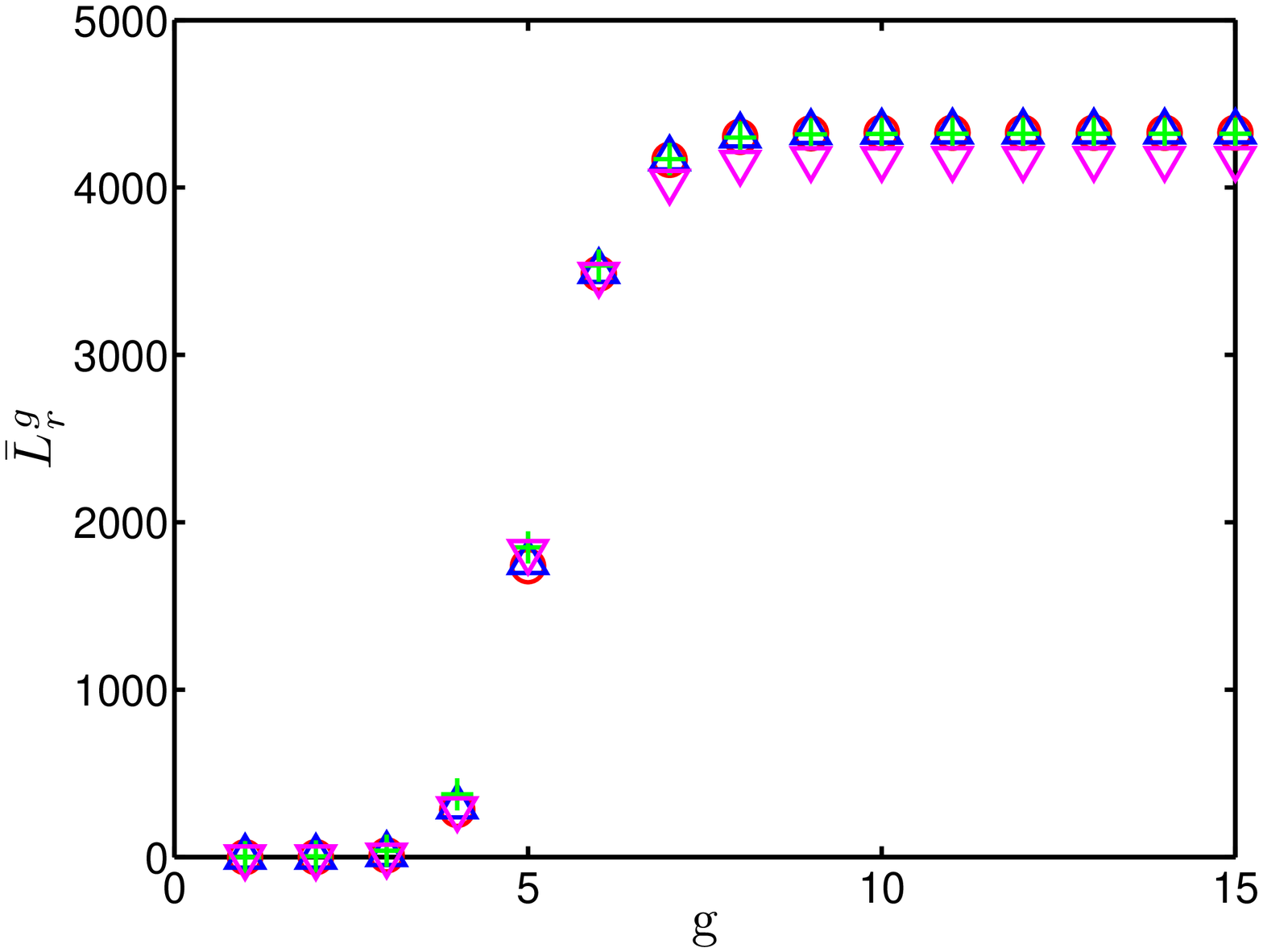} \includegraphics[scale=0.35]{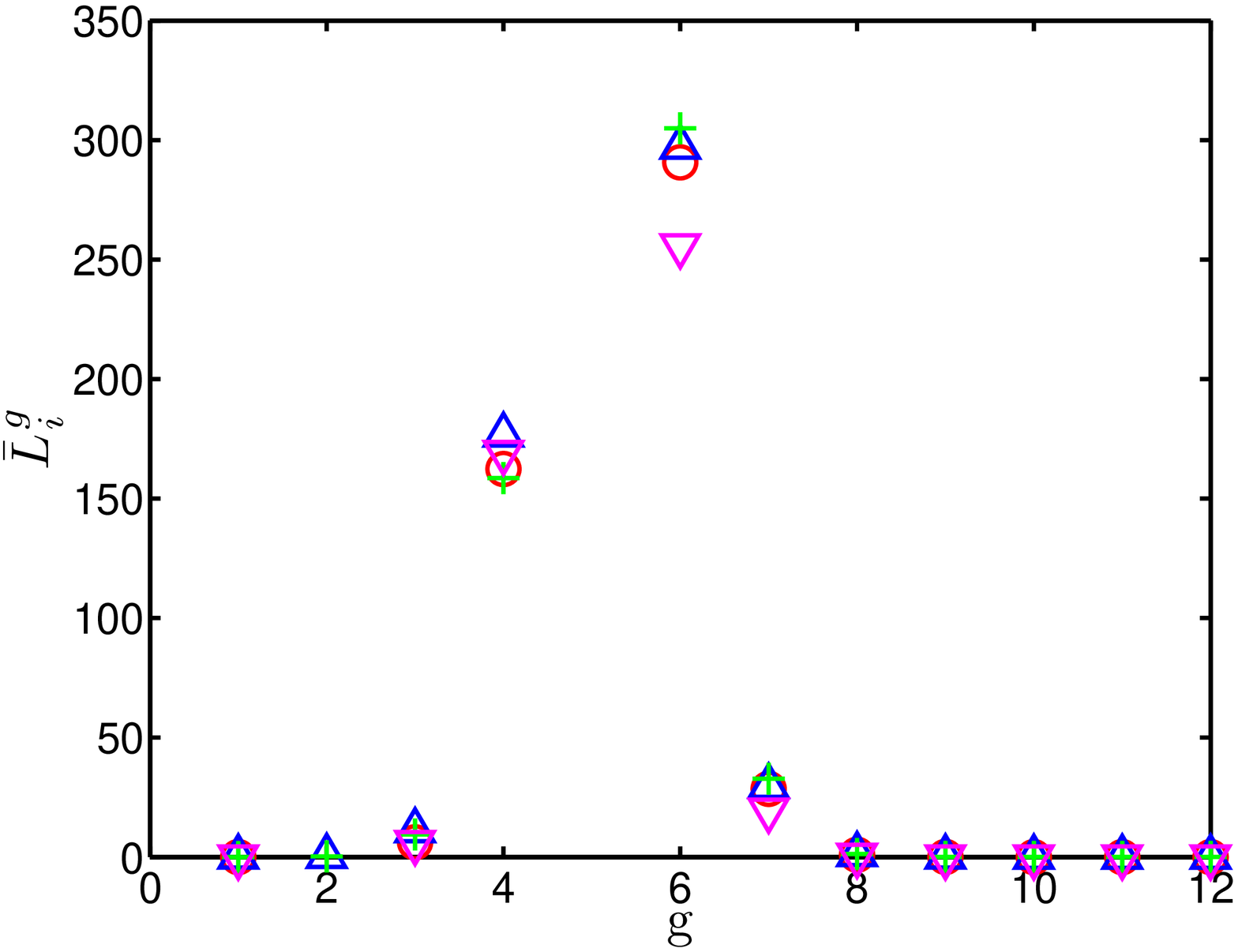}
\caption{The average number of intra-link for the removed (left panel) and
infectious (right panel) classes in terms of generations. The results
are for the exponential distribution, $T=0.3$ and $N=1000$.\label{fig2}}

\end{figure}

\begin{figure}[htb]
\centering{}\includegraphics[scale=0.4]{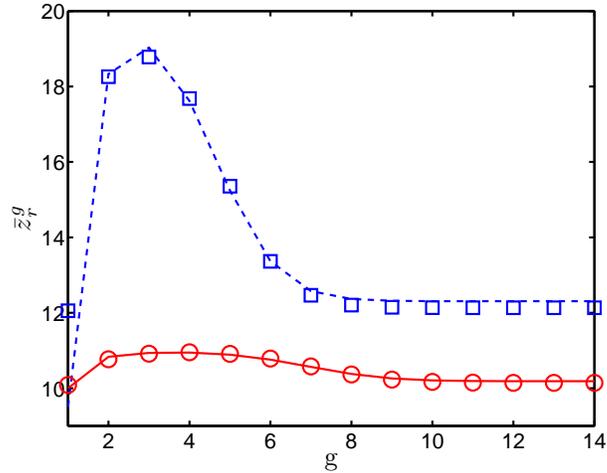} \caption{The average degree of the removed class in terms of generations. The
solid (Poisson) and dashed (exponential) lines represent the analytical
results and the circles (Poisson) and squares ( exponential) represent
the simulation results.\label{fig3}}

\end{figure}

Figure \ref{fig2} shows the same result for an exponential network.
The main difference between the diamonds and the other results comes
from the average degree given by the analytical formula (see Figure
\ref{fig3} and the related discussion). In fact the good agreement
between plus signs, triangles and circles confirms the validity of
equations \eqref{2} and \eqref{3}.

We depict the average degree of the removed class in terms of generations
in Figure \ref{fig3}. The solid (Poisson) and dashed (exponential)
lines present the analytical results and the circles (Poisson) and
squares (exponential) present the simulation results. It is interesting
to note the significant variation of average degree for the exponential
degree distribution. This is mainly due to the width of the degree
distribution. Moreover, the significant difference between the analytical
and simulation results for the exponential degree distribution during
the first generation is due to the exclusion of small outbreaks.

\begin{figure}[htb]
\centering{}\includegraphics[scale=0.35]{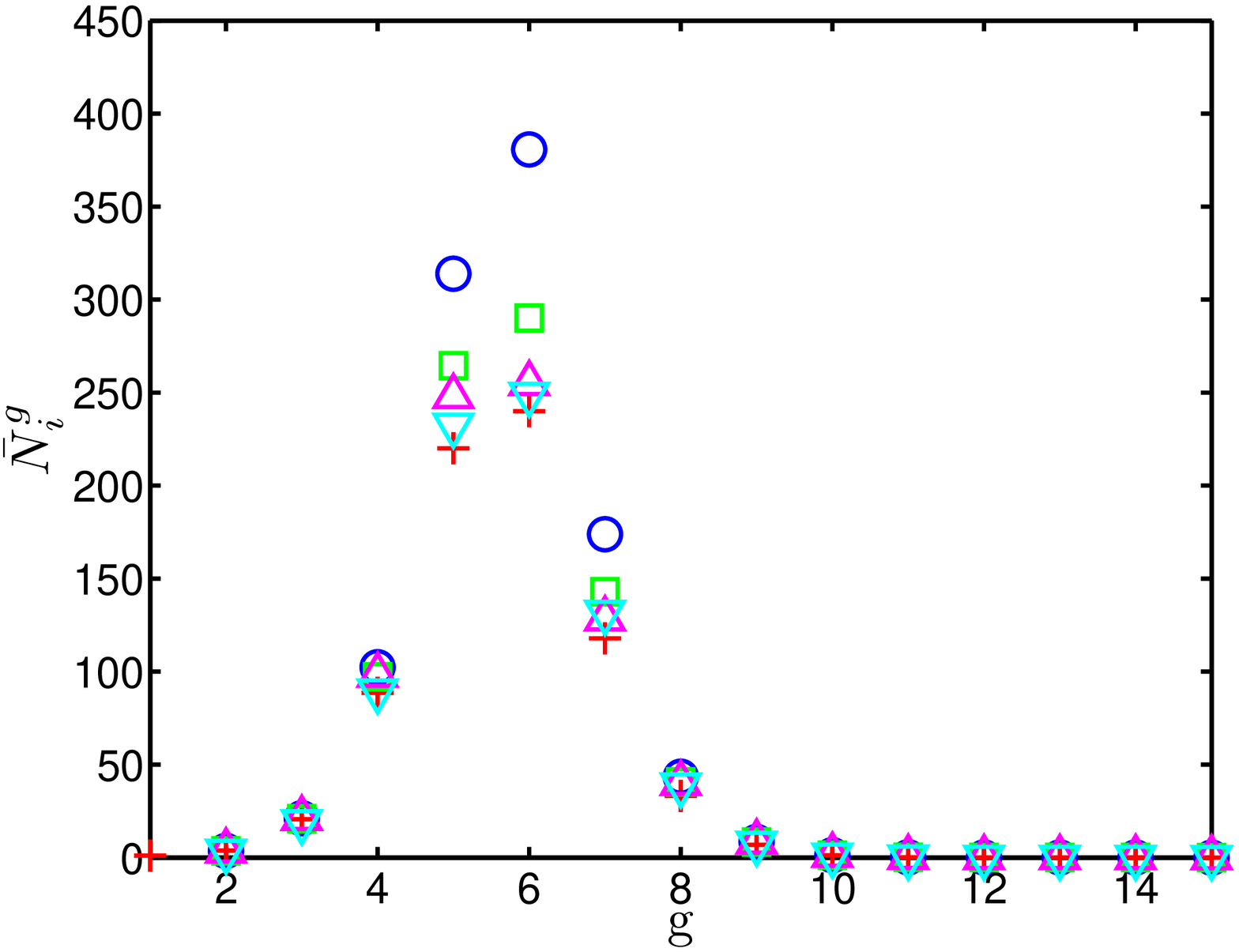} \includegraphics[scale=0.35]{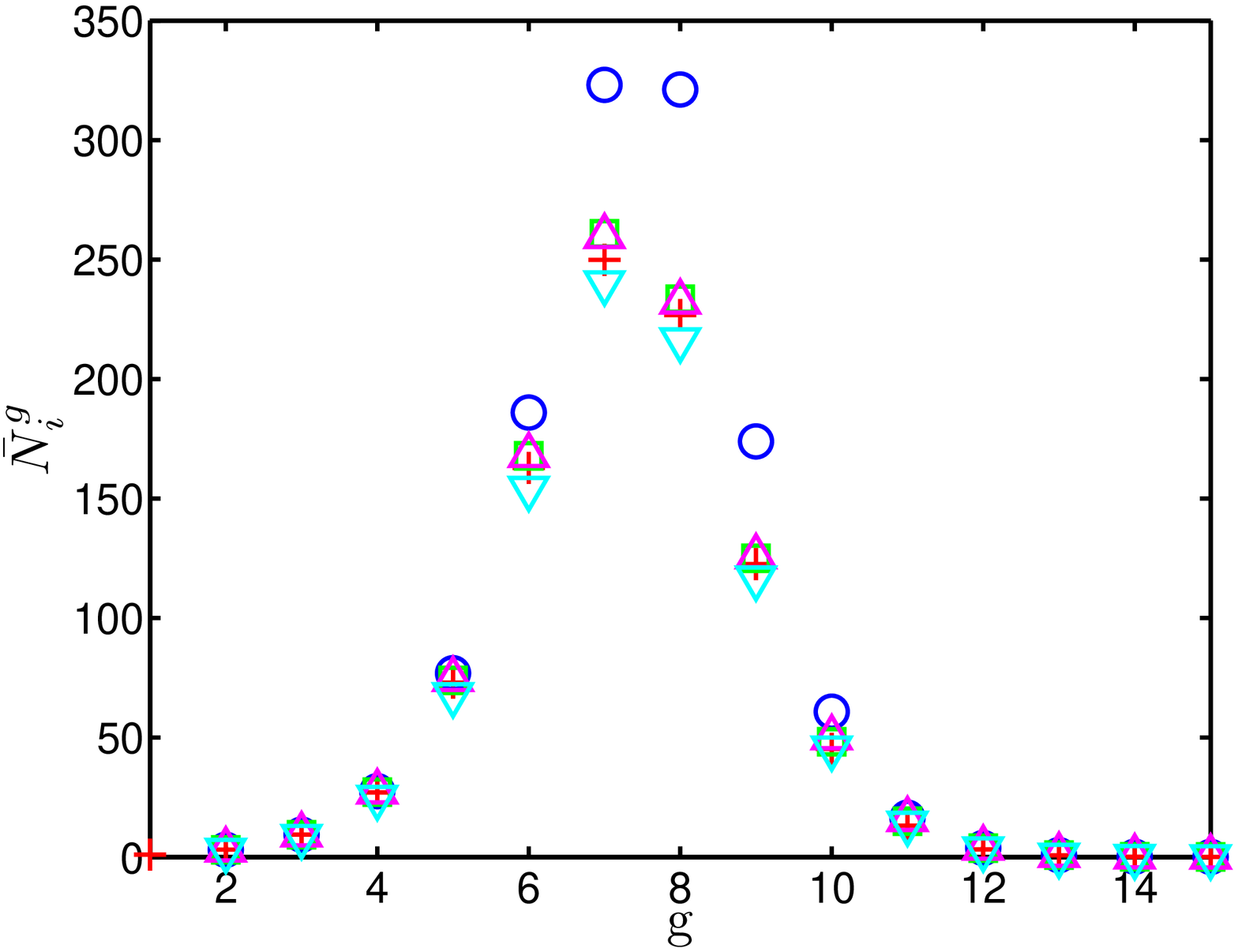}
\caption{The number of new infections and the average number of transmitting
links for exponential (left panel) and Poisson (right panel) for the
same parameters as previous figures. The circles represent the average
number of the transmitting links. The squares, triangles and diamonds
are the number of the new infections calculated by analytical formulas.
The plus signs correspond to simulation results.\label{fig4}}

\end{figure}

We depict in Figure \ref{fig4} the number of new infections and the
average number of transmitting links for exponential (left panel)
and Poisson (right panel) degree distributions for the same parameters
as in previous figures. The circles present the average number of
transmitting links, the squares and plus signs are the number of new
infections calculated by the analytical formula \eqref{4} and obtained
by simulations, respectively, before implementing the corrections
that were derived analytically in equations \eqref{5} and \eqref{6}.
The disagreement between the simulation and the analytical results
for the exponential graph is partly due to the presence of vertices
of very high degree in the network or, more precisely, the asymmetry
of the degree distribution with respect to the mean degree. It is
straightforward to show that the majority of the stubs for an exponential
graph belong to vertices with degree equal to or higher than the average
degree (35\% percent of vertices have 70\% of the stubs for the current
exponential graph). This means that the chance of multi-targeting
a high degree vertex is larger compared to other vertices and this
effect can reduce the average number of new infections compared to
what our analytical formula \eqref{4} predicts. This is taken care
of by the equation \eqref{5} and the result is shown by the triangles.
The change is very small for the Poisson network because of the symmetry
of the degree distribution about the average degree and also because
of the small width of the degree distribution. The inclusion of the
non-transmitting links ($L_{rs}^{g}$ and $\bar{\lambdabar}_{g}$)
can be important when the average degree is low or when the degree
distribution has a heavy tail near the origin. These links in fact
can exclude many vertices by targeting their stubs and reduce the
effective $N_{s}^{g}$ that the transmitting links can target. Our
calculations based on equations \eqref{5} and \eqref{6} are depicted
by diamonds. The effect of equation \eqref{6} might be more important
for lower values of transmissibility when the number of consumed susceptible
vertices due to non-transmitting links is large.

\begin{figure}[htb]
\centering{}\includegraphics[scale=0.35]{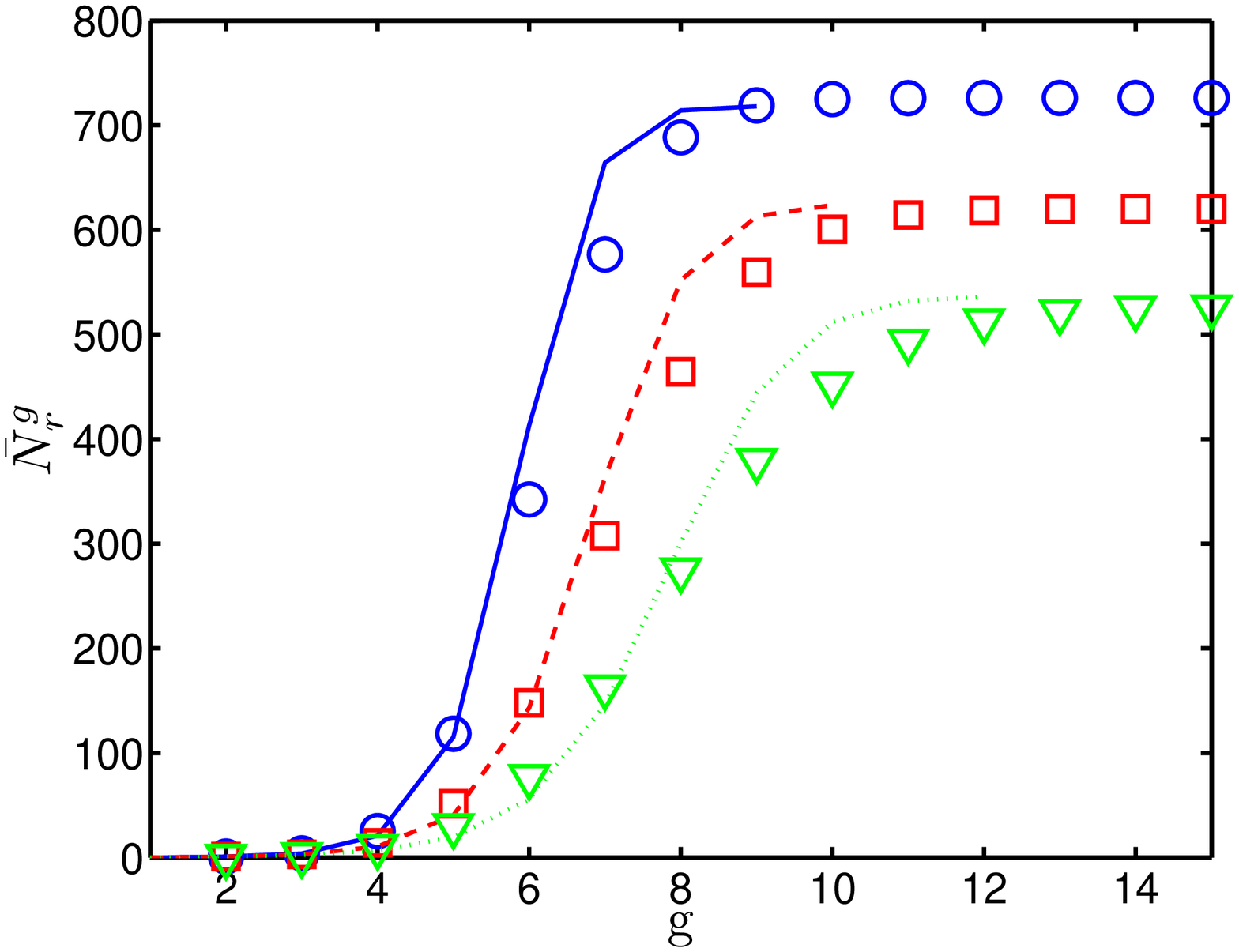} \includegraphics[scale=0.35]{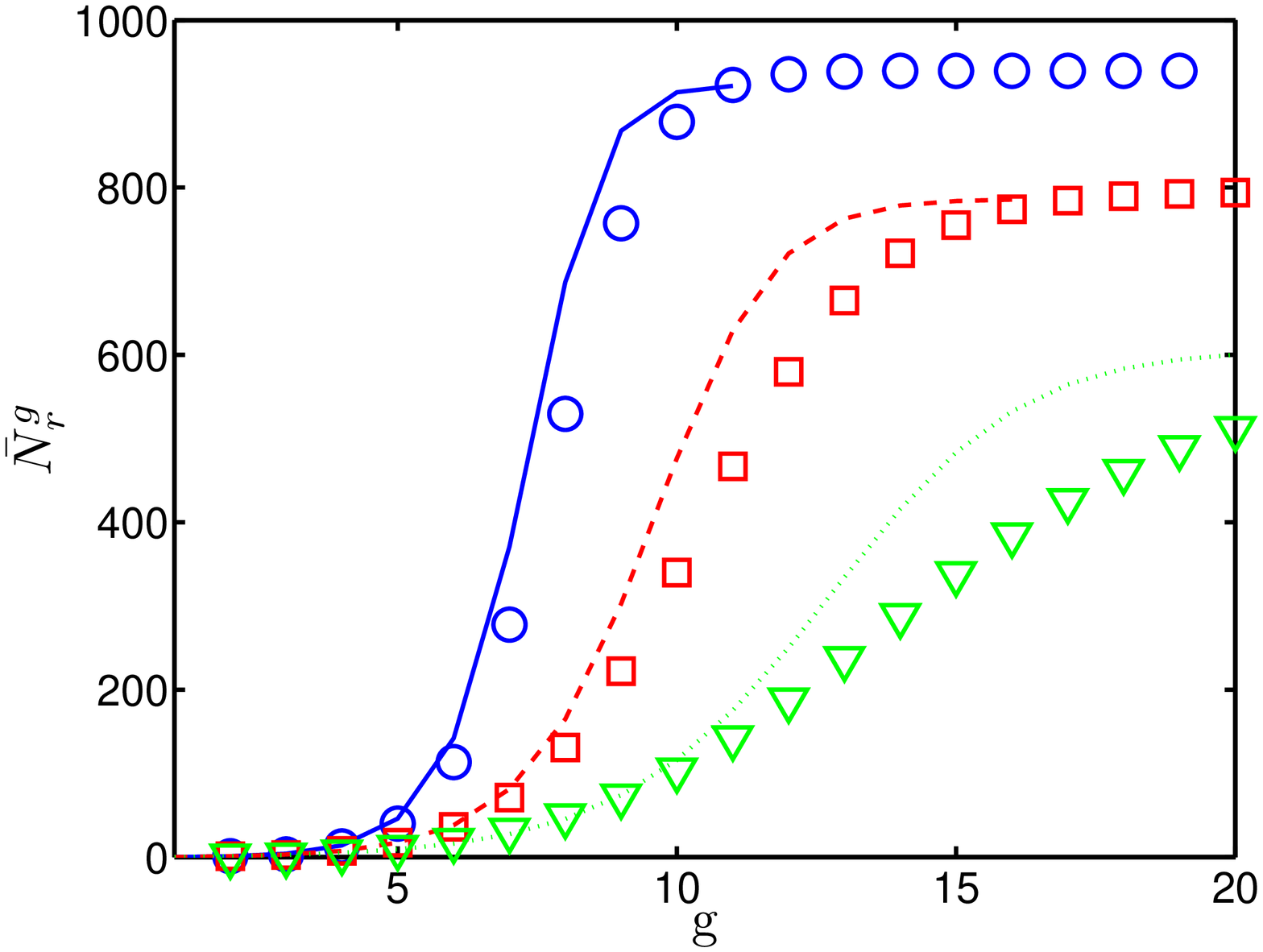}
\caption{The average number of removed vertices as a function of generation
for the exponential (left panel) and Poisson (right panel) at $T=0.15$,
0.2 and 0.3. The lines and points are the analytical and simulation
results respectively.\label{fig5} }

\end{figure}

Now that we have confirmed the validity of the equations introduced
earlier, we can begin solving them together to obtain the average
epidemic size as a function of the generation. We start with one infection
which means, $N_{r}^{0}=0$, $N_{i}^{0}=1$, $N_{s}^{0}=N-1$ and
$\bar{L}_{is}^{0}=z$. Using these initial conditions we can calculate
the number of transmitting links $\bar{\lambda}^{0}$and then $N_{i}^{1}$.
The procedure can easily be continued to the end of the epidemic.
In Figure \ref{fig5} we show the average number of removed vertices
as a function of generation for the exponential (left panel) and Poisson
(right panel) at $T=0.15$, 0.2 and 0.3. The contribution of the small
components is removed from the simulation results as they shift the
mean number of removed vertices to a lower value. This size can be
found simply by looking at the spectrum of component size. The difference
between the analytical and simulation results is partly due to the
initial stage of the simulated epidemics during which the stochastic
behaviour of the outbreaks in different computer runs before entering
to the large-scale epidemic have different time lengths.

\section{Conclusions and Discussion}

We have presented a new approach, which describe the time evolution
of the disease spread on a network. We obtain a set of equations,
which determine the number of different links in each generation.
This step is very important because different links have their own
roles in disease spread. Different types of approximation are discussed
and their validity is tested by simulation. 

This approach can be extended easily to continuous time. Finding the
appropriate dimension of a finite size network, which defines the
number of nth neighbors, is a crucial step in this regard. One can
also extend the approach to evaluate the probability of different
outbreak sizes instead of the average outbreak size. The time evolution
of the degree distribution for different compartments needs more attention.
This requires a serious analysis of the effect of small components
in the network. Knowing the relation between the average degree and
the size of a component is a crucial element of this task.

\end{document}